\documentclass[preprint,reprint,amsmath,showpacs,amssymb,superscriptaddress]{revtex4-1}

\usepackage{graphicx}
\usepackage{dcolumn}
\usepackage{bm}

\begin{document}
\title{Observation of the Kondo Effect in a Spin-$\frac{3}{2}$ Hole Quantum Dot}

\author{O. Klochan}
\email{klochan@phys.unsw.edu.au}
\author{A.P. Micolich}
\author{A.R. Hamilton}
\email{Alex.Hamilton@unsw.edu.au}
\affiliation{School of Physics, University of New South Wales,
Sydney NSW 2052, Australia.}

\author{K. Trunov}
\author{D. Reuter}
\author{A. D.  Wieck}
\affiliation{Angewandte Festk\"{o}rperphysik, Ruhr-Universit\"{a}t Bochum, D-44780 Bochum,
Germany}

\date{\today}

\begin{abstract}

We report the observation of Kondo physics in a spin-$\frac{3}{2}$
hole quantum dot. The dot is formed close to pinch-off in a hole
quantum wire defined in an undoped AlGaAs/GaAs heterostructure. We
clearly observe two distinctive hallmarks of quantum dot Kondo
physics. First, the Zeeman spin-splitting of the zero-bias peak in
the differential conductance is independent of gate voltage. Second,
this splitting is twice as large as the splitting for the lowest
one-dimensional subband. We show that the Zeeman splitting of the
zero-bias peak is highly-anisotropic, and attribute this to the
strong spin-orbit interaction for holes in GaAs.

\end{abstract}

\maketitle

The observation of an unexpected minimum in the low temperature
resistance of metals by de Haas in 1933 was ultimately explained
thirty years later by Kondo as being due to interactions between a
single magnetic impurity and the sea of conduction electrons in a
metal~\cite{deHaasPhysica33, KondoPTP64}. More recently there has
been a resurgence of interest in the Kondo effect, following the
discovery that the conductance of a few electron quantum dot in the
Coulomb blockade regime is enhanced when the dot contains an odd
number of electrons~\cite{KouwenhovenPW01, GoldhaberNat98,
CronenwettSci98}. There is a direct analogy with the Kondo effect in
metals, with the localized electron in the quantum dot acting as a
magnetic impurity that interacts with the two-dimensional sea of
electrons in the source and drain reservoirs.

Studies of the Kondo effect in bulk systems have progressed since
the 1960s, with the focus shifting towards manifestations of Kondo
physics in the strongly correlated electron systems formed in
cuprates and heavy-fermion metals~\cite{StewartRMP01}. More precise
control via improved electrostatic gate design has similarly allowed
progress towards the study of more exotic manifestations of Kondo
phenomena in quantum dots such as the
integer-spin~\cite{SasakiNat00, vanderWielPRL02, GrangerPRB05},
two-impurity~\cite{JeongSci01}, and orbital Kondo
effects~\cite{Jarillo-HerreroNat05}. Thus far all quantum dot Kondo
studies have involved electrons, and GaAs hole quantum dots present
an interesting next step. Holes in GaAs originate from $p$-like
orbitals and behave as spin-$\frac{3}{2}$ particles due to strong
spin-orbit coupling~\cite{WinklerPRL00}. In two- and one-dimensional
systems, the spin-$\frac{3}{2}$ nature of holes leads to remarkable,
highly-anisotropic phenomena~\cite{PapadakisPRL00, DanneauPRL06,
RokhinsonPRL06, KlochanNJP09, ChenNJP10} not observed in electron
systems, and new physics is expected for hole quantum dots
also~\cite{AndlauerPRB09}. Studies of Kondo physics in hole quantum
dots may also provide useful connections to recent studies in bulk
strongly correlated systems~\cite{SuzukiPRB10, SlebarskiPRB10}.

Here we report the observation of the Kondo effect in a GaAs hole
quantum dot. Due to the poor stability of conventional gate-defined
modulation doped structures~\cite{EnsslinNP06} it has not been
possible to define hole quantum dots small enough for studies of
Kondo physics~\cite{GoldhaberNat98}. Instead, we follow the approach
of Sfigakis {\it et al.}~\cite{SfigakisPRL08}, where roughness in
the walls of a wet-etched quantum wire led to formation of an
incidental quantum dot exhibiting Kondo physics as the wire
approached pinch-off. A key advantage to this approach is the
ability to obtain an independent estimate of the effective Land\'{e}
$g$-factor $g^{*}$. Using this we have fabricated a small hole
quantum dot and conclusively demonstrate the ``smoking gun'' for
Kondo physics~\cite{MeirPRL93} -- a splitting of the zero-bias peak
in the differential conductance that opens as $2g^{*}\mu_{B}B$ in
response to an in-plane magnetic field $B$ and is independent of the
gate voltage~\cite{GoldhaberNat98,CronenwettSci98}. In contrast to
electrons, we find that the field splitting of the zero-bias peak is
highly anisotropic.

We used a heterostructure consisting of the following layers grown
on a (100)-oriented substrate: $1~\mu$m undoped GaAs, $160$~nm
undoped AlGaAs barrier, $10$~nm undoped GaAs spacer and a $20$~nm
GaAs cap degenerately doped with carbon for use as a metallic
gate~\cite{ClarkeJAP06, KlochanAPL06}. A (100) heterostructure was
used to avoid the crystallographic asymmetries that plague (311)A
heterostructures~\cite{KlochanNJP09, ChenNJP10}. Ohmic contacts are
made with AuBe alloy annealed at $490^{\circ}$C for $60$~s. Our
devices are remarkably stable, owing to population with holes
electrostatically rather than by ionized modulation
dopants~\cite{ClarkeJAP06, KlochanAPL06}.  A $300$~nm long by
$300$~nm wide quantum wire aligned along the $[01\overline{1}]$
crystallographic direction is fabricated by electron beam
lithography, as shown in the lower left inset to Fig.~1(a). The
quantum wire is defined by shallow etching the doped cap to a depth
$\sim 25$~nm to form three gates -- a central top-gate negatively
biased to $V_{TG}$ to control the hole density and two side-gates
positively biased to $V_{SG}$ to control the electrostatic width of
the wire~\cite{KlochanAPL06, ChenNJP10}. The quantum dot forms as
the wire approaches pinch-off, as discussed in the following
paragraph. All data were obtained at $V_{TG} = -0.67$~V corresponding
to a 2D hole density $p = 2 \times 10^{11}$~cm$^{-2}$ and mobility
$\mu = 450,000$~cm$^{2}$/Vs. We used standard lock-in techniques to
measure the two terminal differential conductance
$G^{\prime}(V_{SD})$ with a variable dc source-drain bias $V_{SD}$
added to a constant $15~\mu$V ac excitation at $5$~Hz. A constant
series resistance of $30.5$~k$\Omega$ was subtracted from all
measurements presented. The experiment was performed in a dilution
refrigerator with a base temperature of $25$~mK, which featured an
{\it in-situ} rotator that enabled the sample to be reoriented with
respect to the applied magnetic field $B$ without the sample
temperature exceeding $200$~mK~\cite{YeohRSI10}.

\begin{figure}
\includegraphics[width = 8.5cm]{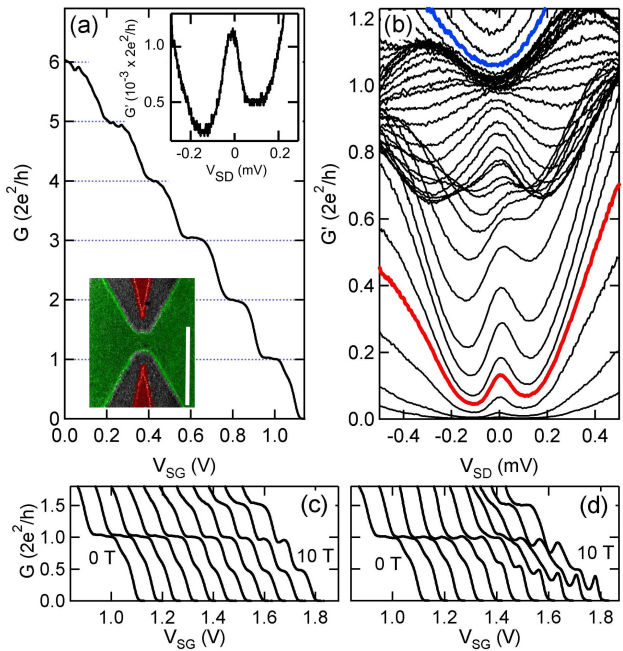}
\caption{\label{fig1} (a) Linear conductance $G$ vs side-gate
voltage $V_{SG}$ for our device, the dot forms at $G < 2e^{2}/h$
(see text). Lower left inset: Scanning electron micrograph of the
device with top-gate (green) and side-gates (red) separated by
etched trenches (grey). The scale bar indicates $1~\mu$m. Upper
right inset: Differential conductance $G^{\prime}$ vs dc
source-drain bias $V_{SD}$ at a side-gate voltage $V_{SG} = 1.115$~V
showing the zero-bias peak at $10^{-3}\times 2e^{2}/h$. (b)
$G^{\prime}$ vs $V_{SD}$ for various $V_{SG}$ starting at $1.1$~V
(bottom) and stepping sequentially by $-5$~mV to $0.9$~V (top). The
red trace ($V_{SG} = 1.085$~V, $G < e^{2}/h$) shows enhanced
conductance centered at $V_{SD} = 0$ while the blue trace ($V_{SG} =
0.91$~V, $G > 2e^{2}/h$) shows the standard parabolic dependence of
$G^{\prime}$ on $V_{SD}$. (c/d) $G$ vs $V_{SG}$ for increasing
in-plane magnetic field (c) parallel to ($B_{\parallel}$) and (d)
perpendicular to ($B_{\perp}$) the wire. Traces obtained with a $+
1$~T increment and offset to the right by $+ 0.07$~V for clarity.}
\end{figure}

Figure~1(a) shows the linear conductance $G = G^{\prime}(V_{SD} =
0)$ versus $V_{SG}$ with the six quantized conductance plateaus
confirming ballistic transport through the device. The quantum dot
forms at $G < 2e^{2}/h$ due to a combination of microscopic
deviations in confining potential due to etch roughness in the
gates~\cite{SfigakisPRL08} and self-consistent electrostatic
effects~\cite{YoonPRL07}. The presence of a bound-state in this
system is revealed by the evolution of $G$ versus $V_{SG}$ with
increasing in-plane magnetic field aligned parallel $B_{\parallel}$
[Fig.~1(c)] and perpendicular $B_{\perp}$ [Fig.~1(d)] to the wire at
$G < 2e^{2}/h$. In both cases, plateaus at $e^{2}/h$ and $3e^{2}/h$
emerge, indicating the onset of spin-splitting~\cite{PatelPRB91},
accompanied by sharp resonances signalling formation of a
bound-state within the wire~\cite{McEuenSurfSci90, LiangPRL98,
SfigakisPRL08, KomijaniEPL10}. We will show later in Fig.~2(a) that
this bound-state is not an impurity effect, as it is robust to
gate-induced lateral shifting of the 1D channel~\cite{GlazmanSST91}.
Coupling of the magnetic field to orbital motion~\cite{QuayNP10} may
be responsible for differences in resonant structure between
Figs.~1(c/d), as any random disorder potential should be constant
given both orientations were measured during a single cooldown.

A study of the zero-bias peak (ZBP) in the differential conductance
provides additional evidence for quantum dot formation. Figure~1(b)
shows $G^{\prime}$ versus $V_{SD}$ at a range of $V_{SG}$ spanning
$0 < G < 1.2 \times 2e^{2}/h$, with two traces highlighted. Under a
single particle picture, $G^{\prime}$ should depend quadratically on
$V_{SD}$ with a minimum at $V_{SD} = 0$~\cite{MartinMorenoJPCM92},
as observed for the blue trace at $G > 2e^{2}/h$ in Fig.~1(b). In
contrast at $G < 2e^{2}/h$, $G^{\prime}$ shows a pronounced peak at
$V_{SD} = 0$ superimposed upon a parabolic background [red trace,
Fig.~1(b)], known as the zero-bias peak. We observe the ZBP at
conductances as low as $10^{-3} \times 2e^{2}/h$, consistent with
previous work~\cite{SfigakisPRL08, SarkozyPRB09, RenPRB10}. This is
demonstrated upper right inset to Fig.~1(a).

\begin{figure}
\includegraphics[width = 8.5cm]{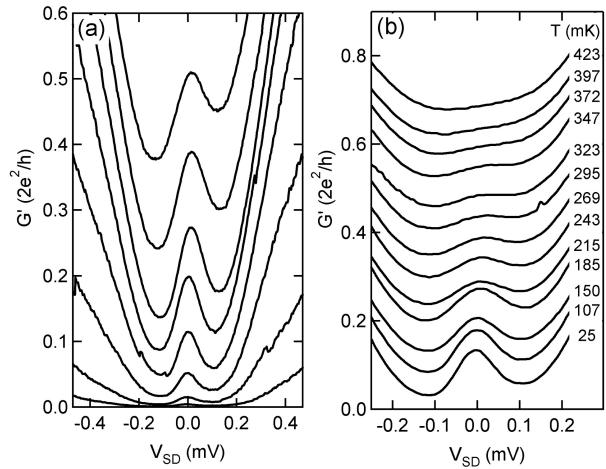}
\caption{\label{fig2} (a) Asymmetric bias study of $G^{\prime}$ vs
$V_{SD}$ at various $V_{SG}$. Each trace directly corresponds to one
in Fig.~1(b), with the two side-gate voltages set such that they
differ by $0.5$~V but their average equals the $V_{SG}$ in
Fig.~1(b). The lowest trace has $V_{SG}^{1} = 0.85$~V and
$V_{SG}^{2} = 1.35$~V, which average to $1.1$~V to match $V_{SG}$
for the lowest trace in Fig.~1(b). Moving upwards, both side-gates
are sequentially incremented by $-5$~mV for each trace, the
uppermost trace has an average $V_{SG} = 1.065$~V. (b) $G^{\prime}$
vs $V_{SD}$ at fixed $V_{SG} = 1.085$~V for different temperatures
$T$. Traces are sequentially offset by $+ 0.05 \times 2e^{2}/h$ from
bottom.}
\end{figure}

To demonstrate that the ZBP is robust and not due to random
disorder~\cite{ChenPRB09}, we have studied the ZBP as the wire is
shifted laterally. We do this by repeating each measurement in
Fig.~1(b) with a voltage offset of $0.25$~V added to side-gate 1 and
subtracted from side-gate 2, such that the average bias is
maintained to facilitate direct comparison. This results in a
lateral shift of the wire by $\approx$60~nm~\cite{GlazmanSST91}, and
gives the data shown in Fig.~2(a) For traces below $e^{2}/h$ there
is almost no change in the ZBP compared to the data from the
unshifted channel in Fig.~1(b), confirming that the ZBP is not
disorder-induced. Equivalent data was obtained when we shifted the
channel in the opposite direction (not shown). To further check the
consistency of our ZBP with the known Kondo physics of
electron quantum dots, in Fig.~2(b) we show the evolution of the red
trace from Fig.~1(b) with temperature $T$. The peak widens and
decreases in amplitude with increasing $T$, consistent with previous
studies~\cite{CronenwettSci98}.

\begin{figure}
\includegraphics[width = 8cm]{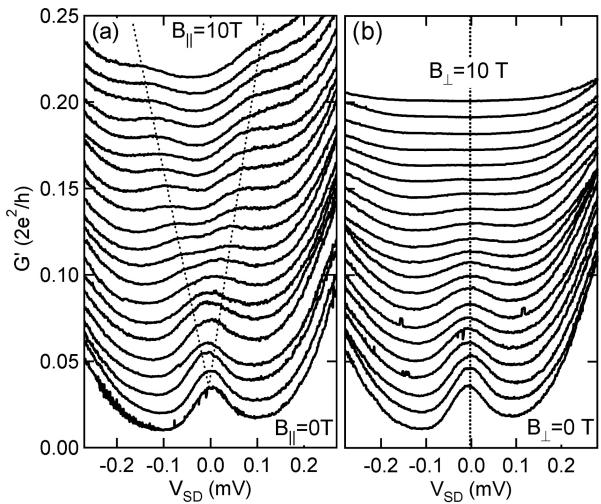}
\caption{\label{ZBA-1D} $G^{\prime}$ vs $V_{SD}$ at $V_{SG} =
1.105$~V measured for in-plane magnetic fields ranging from $0$ to
$10$~T oriented (a) parallel to ($B_{\parallel}$) and (b)
perpendicular to ($B_{\perp}$) the wire. Traces obtained with a
$+0.5$~T increment and sequential vertical offset of $+ 0.01 \times
2e^{2}/h$ from the bottom.}
\end{figure}

We now focus on the magnetic field dependence of the zero-bias peak
looking for: a) the $g^{*}$ anisotropy characteristic of
spin-$\frac{3}{2}$ holes, and b) the distinctive $2g^{*}\mu_{B}B$
peak splitting of Kondo physics. We begin by examining the evolution
of the ZBP with $B_{\parallel}$ in Fig.~3(a). Initially no splitting
is resolved and the only change is a widening of the zero-bias peak.
However, at $B_{\parallel} \approx 4$~T two peaks become resolved,
and these separate in $V_{SD}$ as $B_{\parallel}$ is increased
further. The behavior is very different with an in-plane field
$B_{\perp}$ applied perpendicular to the wire [Fig.~3(b)]. Here the
peak shows no splitting even at the highest field $B_{\perp} =
10$~T; instead the peak is gradually reduced in amplitude and
ultimately suppressed entirely. This anisotropic behavior matches
the underlying $g^{*}$ anisotropy of the 1D wire~\cite{ChenNJP10} in
which the dot resides. We return to this anisotropy in the final
discussion, and now continue with quantitative analysis of the peak
splitting with $B_{\parallel}$.

\begin{figure}
\includegraphics[width = 7cm]{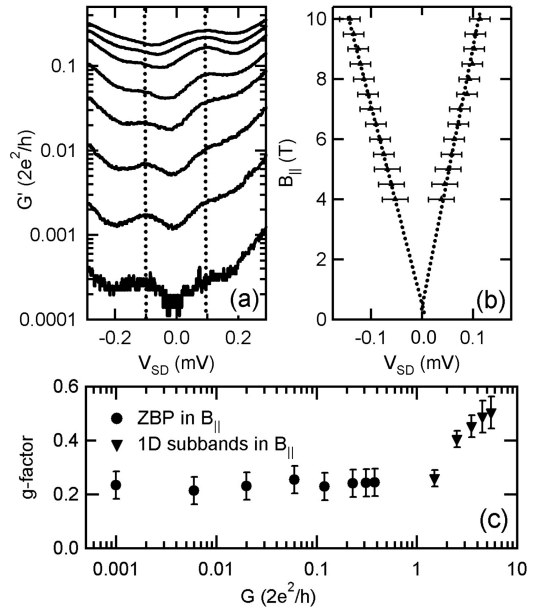}
\caption{\label{gfactors} (a) Plot of $G^{\prime}$ vs $V_{SD}$ at various $V_{SG}$
at fixed $B_{\parallel} = 8$~T demonstrating the gate voltage
independence of the splitting.  (b) Location of the spin-split zero-bias
peaks in $V_{SD}$ ($x$-axis) vs $B_{\parallel}$ for the data from
Fig.~3(a). (c) Measured $g$-factor $g^{*}$ for
$B_{\parallel}$ vs $G$ for the zero-bias peak (solid circles) at $G
< 2e^{2}/h$ and 1D subbands (solid triangles) at $G > 2e^{2}/h$.}
\end{figure}

As pointed out by Cronenwett {\it et al.} in Ref.
~\cite{CronenwettSci98}, the most distinct sign of the quantum dot
Kondo effect is a gate voltage independent ZBP split by
$2g^{*}\mu_{B}B$. Figure 4(a) shows $G^{\prime}$ versus $V_{SD}$ at
various $V_{SD}$ at $B_{\parallel} = 8$~T. The two vertical dotted
lines in Fig.~4(a) pass through the field-split zero-bias peaks over
more than three orders of magnitude in conductance, demonstrating
the gate voltage independence of the peak splitting. Turning to the
splitting as a function of field, in Fig.~4(b) we plot the peak
location in $V_{SD}$ for the traces in Fig.~3(a) where two peaks can
be clearly resolved against $B_{\parallel}$. The peak locations are
determined by eye and the error bars are estimated knowing that the
ZBP sits on a parabolic background with a slight linear asymmetry
from the way that $V_{SD}$ is applied in the measurement circuit.
The ZBP clearly splits linearly with $B_{\parallel}$ in Fig.~4(b),
giving $g^{*}_{\text{ZBP}} = 0.23 \pm 0.05$ if we assume a splitting
$eV_{SD} = 2g^{*}_{\text{ZBP}}\mu_{B}B$. The data will justify this
assumption below. We repeated this analysis at eight different
conductances between $10^{-3}$ and $0.5 \times 2e^{2}/h$ giving the
solid circles plotted in Fig.~4(c). The error bars are obtained from
a regression analysis of linear fits such as that in Fig.~4(b). The
$g^{*}_{\text{ZBP}}$ values obtained are constant over three orders
of magnitude in $G$, in agreement with Fig.~4(a), and give an
average $g^{*}_{\text{ZBP}} = 0.236 \pm 0.012$.

A natural question is: How do we know that the peak splitting is
given by $2g^{*}\mu_{B}B$ rather than $g^{*}\mu_{B}B$, which would
increase the $g^{*}$ extracted from the data by a factor of two? The
key advantage of our device is that we can use 1D subband
spectroscopy~\cite{PatelPRB91} to independently measure $g^{*}$ in
the limit where $G$ approaches $2e^{2}/h$ from above. This allows us
to corroborate our measurement of $g^{*}_{\text{ZBP}}$. We measure
$g^{*}_{\text{1D}}$ for the first five 1D subbands using the method
in Ref.~\cite{ChenNJP10} and the usual Zeeman expression
$g^{*}_{\text{1D}}\mu_{B}B$ for 1D subbands~\cite{PatelPRB91}. The resulting $g^{*}_{1D}$ values, plotted as
solid triangles in Fig.~4(c), decrease linearly as $G$ approaches
$2e^{2}/h$. This linear decrease is distinctive of holes in (100)
heterostructures~\cite{ChenNJP10}. However, the most significant
aspect is that the $g^{*}_{\text{1D}} = 0.25 \pm 0.03$ obtained for
the lowest 1D subband assuming the 1D splitting goes as
$g^{*}\mu_{B}B$ is in excellent agreement with $g^{*}_{\text{ZBP}}$
obtained assuming the ZBP splitting goes as $2g^{*}\mu_{B}B$. This
is ``smoking gun'' evidence confirming our observation of Kondo
physics in a hole quantum dot~\cite{MeirPRL93, GoldhaberNat98,
CronenwettSci98}.

We conclude by discussing some key implications of our findings. The
magnitude and anisotropy of $g^{*}_{\text{ZBP}}$ closely matches
that of $g^{*}_{\text{1D}}$ for the lowest 1D subband. This suggests
that $g^{*}_{\text{ZBP}}$ is set by the prevailing $g^{*}$ of the
environment hosting the dot. It agrees with quantum
dots~\cite{CronenwettSci98}, where the splitting of the ZBP gives
the same $g^{*}$ as bulk GaAs, and with carbon
nanotubes~\cite{NygardNat00}. The fact that a spin-$\frac{3}{2}$
system produces no radical change in the observed Kondo physics is
interesting, as it implies that the process only relies on the
presence of a doubly-degenerate quantum dot level to mediate
transport between the reservoirs, and not its precise nature/spin.
This is in accordance with recent studies of more exotic
manifestations of Kondo physics in quantum dots~\cite{SasakiNat00,
vanderWielPRL02, GrangerPRB05, JeongSci01, Jarillo-HerreroNat05}.
However, a spin-$\frac{3}{2}$ system may ultimately present more
subtle changes, for example, confinement-induced
mixing~\cite{ZulickePSSC06} between heavy-hole and light-hole
subbands (i.e., states with total angular momentum quantum numbers
$m_{j} = \pm \frac{3}{2}$ and $\pm \frac{1}{2}$ respectively) may
alter the relevant scales in the problem. Further studies in this
direction would be useful, including both theoretical work and
measurements from improved device geometries. Finally, we comment
briefly on the bearing of our results on studies of the $0.7$
anomaly~\cite{ThomasPRL96} in 1D systems. Although we observe a
plateau-like feature near $0.7 \times 2e^{2}/h$ in our device (see
Figs.~1(a,c,d)), the presence of the resonant structure in the
linear conductance precludes any direct and definitive link between
the $0.7$ anomaly and the behavior we observe for our zero-bias
peak. We emphasize that the gate-voltage independent Zeeman
splitting of our ZBP points conclusively to the quantum
dot Kondo effect, in contrast with the gate-voltage dependent
zero-bias anomaly (ZBA) splitting observed in undoped quantum wires
by Sarkozy {\it et al.}~\cite{SarkozyPRB09}. The characteristics of
our zero-bias peak are very different to those of the ZBA in quantum
wires. The two effects clearly have a different origin, which agrees
with the suggestion by Sarkozy {\it et al.}~\cite{SarkozyPRB09} that
the ZBA is a fundamental property of quantum wires, and suggests
that it may involve processes beyond Kondo physics alone.

This work was funded by the Australian Research Council (Grant Nos.
DP0772946, DP0986730, FT0990285). We thank U. Z\"{u}licke for
helpful discussions, L.A. Yeoh and A. Srinivasan for development of
the low temperature rotator, and J. Cochrane for technical support.

\end{document}